\documentclass{ws-procs975x65}

\begin{document}

\title{Ultra High Energy Cosmic Rays: the present position and 
the need for mass composition measurements*}

\author{A A Watson}

\address{School of Physics and Astronomy\\
University of Leeds\\
Leeds LS2 9JT, UK}

\maketitle

\abstracts{ The present situation with regard to experimental data 
on ultra high-energy cosmic rays is briefly reviewed.  Whilst detailed 
knowledge of the shape of the energy spectrum is still lacking, it is 
clear that events above $10^{20}$ eV do exist.  Evidence for 
clustering of the directions of some of the highest energy events remains 
controversial.  Clearly, more data are needed and these will 
come from the southern branch of the Pierre Auger Observatory in the 
next few years.  What is evident is that our knowledge of the mass 
composition of cosmic rays is deficient at all energies above $10^{18}$ eV. 
It must be improved if we are to discover the origin of the highest energy cosmic rays.  The major part of the paper is concerned with this problem: it 
is argued that there is no compelling evidence to support 
the common assumption that cosmic rays of the highest energies are protons.}

\footnotetext{*based on invited
talk at `Thinking, Observing and Mining the Universe', Sorrento
Italy: 22 -- 27 September 2003}

\section{Motivation for Studying the Highest Energy 
Cosmic Rays}  
Efforts to discover the origin of the highest energy 
cosmic rays have been on going for many years.  Since the recognition 
in 1966, by Greisen and by Zatsepin and Kuzmin, that protons with energies 
above $4 \times 10^{19}$ eV would 
interact with the cosmic microwave radiation, there has been 
great interest in measuring the spectrum, arrival direction distribution 
and mass composition of ultra high-energy cosmic rays (UHECR).  
UHECR may be defined as those cosmic rays having energies above 
$10^{19}$ eV.  Specifically, it was pointed out that if the sources 
of the highest energy protons were universally distributed, then there 
should be a sharp steepening of the energy spectrum in the range 
from 4 to $10 \times 10^{19}$ eV.  This 
predicted feature has become known as the GZK cut-off.  If the UHECR 
were mainly Fe nuclei then there would also be a steepening of the 
spectrum.  However, it is harder to predict this feature 
accurately as the relevant diffuse infrared photon field is poorly
known: the steepening is expected to set in at higher energy.

Early instruments built to study this energy region (Volcano 
Ranch, Haverah Park, SUGAR and 
Yakutsk), were designed before the 1966 predictions
and when the flux above $10^{19}$ eV was poorly known.  
Although of relatively small area ($\sim 10$ km$^{2}$) sufficient 
exposure was accumulated to measure the rate of cosmic 
rays above $10^{19}$ eV accurately and to give the first 
indications that there might be cosmic rays with energies 
above $10^{20}$ eV.  
No convincing evidence of anisotropies above $10^{19}$ eV was 
established.  It also became accepted 
that the problem of acceleration of protons and nuclei to such 
energies in known astrophysical sources is a major one.  The projects 
that followed the pioneering ones also gave 
indications of trans-GZK particles but by the early 1990s it was 
apparent that even areas of 100 km$^{2}$, operated for many years, 
could not measure the properties of UHECR with adequate detail.  
Accordingley, work has started on a 3000 km$^{2}$ detector,
the Pierre Auger Observatory.

\section{ The present observational situation}
\begin{figure}[htb]
\centerline{\epsfxsize=4.0in\epsfbox{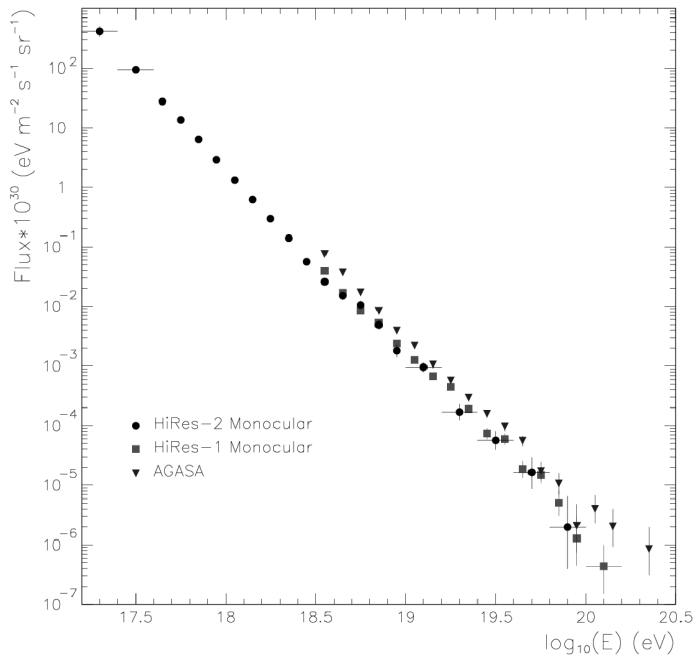}}
\caption{The energy spectra as reported by the AGASA$^{1}$ and 
HiRes$^{2}$ groups.  This clear presentation of the spectra is 
due to D Bergman (University of Columbia). \label{fig1}}
\end{figure}

During the planning and construction of the Pierre Auger 
Observatory, observations continued with the Japanese array (AGASA) 
of plastic scintillators and the two fluorescence detectors 
(HiRes and Fly's Eye) of the University of Utah.  The Japanese 
detector is an array of 111 x 2.2 m$^{2}$ plastic scintillators 
spread over 100 km$^{2}$.  It will cease 
operation in December 2003 when an exposure of about 1600 km$^{2}$ sr 
years will have been made.  
Eleven events with energies above $10^{20}$ eV have been 
reported \cite{tak03}.  Spectra derived from arrays of 
particle detectors suffer from the difficulty that the energy of 
each primary cosmic ray must be inferred using models of particle 
physics interactions at energies well beyond those of present, or envisaged, 
accelerators.  Thus, there is a systematic error in these energy 
assignments that is, inherently, unknowable.  By contrast, the fluorescence method uses the scintillation light produced in the 
atmosphere by the secondary shower cascade and permits a calorimetric 
estimate of the energy in a manner familiar from accelerator 
experiments, although there are difficulties associated with the 
variable transmission properties of the atmosphere and with accurate
knowledge of fluorescence yield.  These instruments have also 
seen events with energies above $10^{20}$ eV but at a rate lower 
than that claimed by the Japanese group.  Nevertheless, the 
highest energy event ever recorded ($3 \times 10^{20}$ eV) was detected 
by the Fly's Eye instrument and it is clear that 
there are cosmic rays above $10^{20}$ eV seen with 
both techniques and that the rate of such events is of 
order 1 per km$^{2}$ per steradian per century.  A useful 
summary of the experimental situation is shown in figure~\ref{fig1}. 

Many questions remain about the detailed shape of the spectrum.   
The HiRes and AGASA spectra could be reconciled if 
the energy scale of one or other was adjusted by 30\%.  Moreover, the possibility that there are 
uncertainties in the flux measurements should not be overlooked.  At 
the lower end of the AGASA spectrum the aperture is 
changing quite rapidly with energy \cite{tak03} and uncertainties in the 
function that describes the fall-off 
of signal with distance may lead to uncertainties in the 
aperture determination.  At the highest energies, the AGASA aperture, 
limited by requiring that shower cores fall inside the array area, 
is known precisely.  By contrast, with fluorescence 
detectors, the aperture continues to grow with energy.  There is 
considerable uncertainly about the HiRes aperture, even in the 
case of stereo operation \cite{hir03}.  Further data are expected 
from the HiRes 
group and, in particular, from their period of stereo operation.

The situation concerning the arrival direction distribution of 
UHECR is not clear-cut either.  For some time the AGASA 
group \cite{aga03} have reported clustering on an angular scale 
of 2.5$^{\circ}$, from a data set of 59 events above 
$4 \times 10^{19}$ eV.  The clusters are claimed to occur much 
more frequently than expected by chance with an estimate of $10^{-4}$ 
given for the chance probability.  A preliminary search of the HiRes 
data \cite{hin03} has not revealed clusters with the same frequency 
as claimed by AGASA.

Recently, Finley and Westerhoff \cite{fin03} have presented an 
analysis using the directions of 72 events recently released by 
the AGASA group.  They have taken the 30 events described 
in \cite{hay99} as the trial data set and used the additional 42 events 
to search for pairs, adopting the criteria established by the 
AGASA group.  Two pairs were found with a probability of 19\% of 
occurring by chance. 

It is clear that only further data will resolve the controversies 
over the energy spectrum and over the clusters in arrival 
direction.  The AGASA array will close at the end of 2003 when it will 
have achieved an exposure of $\sim$ 1600 km$^{2}$ sr years.  
The HiRes instrument is expected to take data for another few 
years.  The Pierre Auger Observatory has been designed to clarify 
the situation.  It makes use of the ground array technique 
and of the fluorescence technique in what has become known as a 
`hybrid detector'.  When completed in late 2005, it will cover 
3000 km$^{2}$ with $1600 \times 10$ m$^{2} \times 1.2$ m deep 
water-Cherenkov detectors on a 1.5 km hexagonal grid.  
These detectors will be overlooked by four fluorescence 
detectors constructed on prominences at the edge of the area.  An 
engineering array has operated for two years and 
all systems have performed within the design 
specifications \cite{pie03}.  While this observatory 
will mainly survey the Southern sky, it is expected to give a guide as 
to which of the spectra so far reported is correct and of 
the reality of clustering: the exposure made by the end of 2004 
is expected to be comparable to that of AGASA.  The immediate prospect, therefore, is for science data to be reported in mid-2005. 
This Southern part of the observatory is seen as the first of the two
that are needed to provide full sky coverage. 

\section{Interpretation of the existing data}
Many attempts have been made to explain the particles that 
exist beyond the GZK cut-off.  If these are protons, the existence 
of such UHECR is seen as an enigma.  They must come from nearby 
(at $10^{20}$ eV about 50\% are expected from within 20 Mpc) and, 
adopting an extragalactic field of a few nanogauss, point sources would be expected to be detectable.  However, none are seen and a wide variety of 
explanations has been offered.  Amongst the many mechanisms proposed are the 
decay of topological defects or other massive relics of the big bang.  
Even more exotic is the suggestion of a violation of Lorentz invariance 
in such a manner that the energy-loss 
mechanism against the CMB is not effective, though acceleration
remains an issue.  If the primaries were 
iron nuclei then the situation would be slightly easier to understand.  
The higher charge would mean that acceleration 
could occur more readily and that 
bending, even in a weak magnetic field, would obscure 
the source directions.  Without data on the mass composition it will 
be hard to draw conclusions about the origin of the particles, even 
when the spectral and clustering issues are clarified.

\section{The mass of UHECR} 
Our knowledge about the mass of primary cosmic rays at energies above 
$10^{17}$ eV is rudimentary.  Different methods of measuring the 
mass give different answers and the conclusions are usually 
dependent upon the model calculations that are assumed.  Results from 
some of the techniques that have been used in attempts to 
assess the mass composition are now described and the conclusions 
drawn reviewed.  Some of these techniques are applicable to the 
Pierre Auger Observatory.

\subsection{The Elongation Rate} 
\begin{figure}[htb]
\centerline{\epsfxsize=5.0in\epsfbox{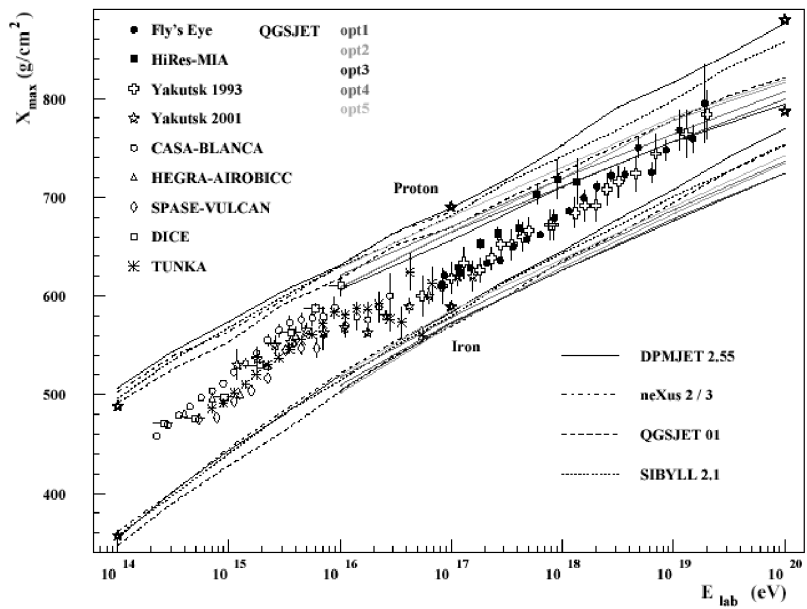}}
\caption{The depth of maximum, as predicted using various 
models, compared with measurements.  
The predictions of the five modifications of QGSJET$^{9}$ from which this 
diagram is taken, 
lie below the dashed line that indicates the predictions 
of QGSJET01. \label{fig2}}
\end{figure}

The elongation rate describes the rate of change of 
depth of shower maximum with primary energy.  The term 
was introduced by Linsley \cite{lin77} and, although his 
original conclusions have been largely superseded by the results of 
Monte Carlo studies, the concept is useful 
for organising data. 
A summary of measurements of the depth of maximum 
together with predictions from a variety of model 
calculations \cite{zha03} is in figure~\ref{fig2}.  It is clear 
that if certain models 
are correct that one might infer that primaries above $10^{19}$ eV 
are dominantly protons; others suggest a mixed composition.  The QGSJET 
set of models (basic QGSJET01 and the 5 options 
discussed in \cite{zha03}) and the Sibyll 2.1 model force contrary 
conclusions.  

\subsection{Fluctuations in Depth of Maximum} 
\begin{figure}[htb]
\centerline{\epsfxsize=4.0in\epsfbox{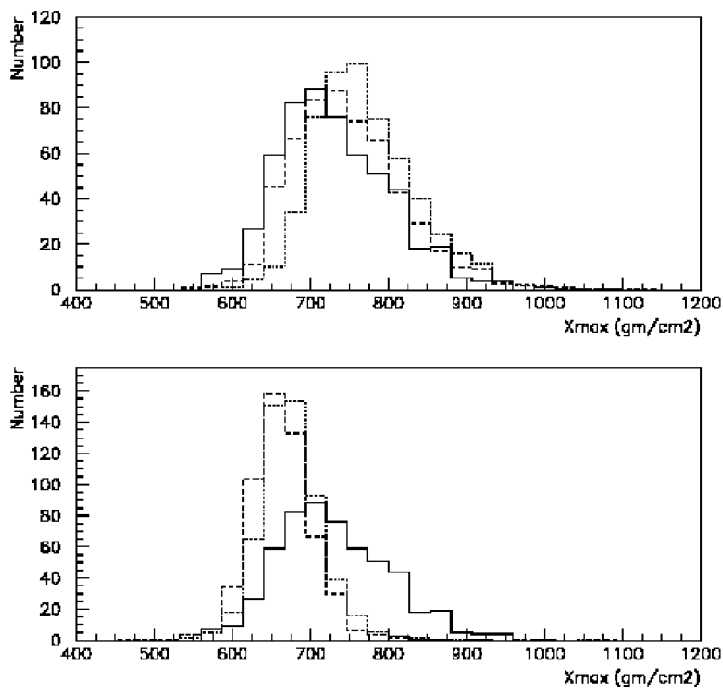}}
\caption{The HiRes data $^{10}$ on X$_{max}$ for $> 10^{18}$eV.  
The solid lines in the two figures are 
the experimental data.  The upper figure shows predictions for 
proton primaries for the QGSJET and Sibyll models.  
Predictions for iron primaries are shown in the lower figure. \label{fig3}}
\end{figure}

A way to break this degeneracy has long been seen in the 
magnitude of fluctuations of the position of depth of maximum.  If a group of showers with a narrow range of energies is selected 
then fluctuations about the 
mean X$_{max}$ would be expected to be larger for protons than for 
iron nuclei.  A recent study of this, as reported by the HiRes group 
\cite{tsu03}, is shown in figure~\ref{fig3}.  These data are for 
553 events $> 10^{18}$ eV.  
It is argued that the fluctuations are so large that a large 
fraction of protons is indicated.  However, the HiRes data have 
been analysed assuming a standard US atmosphere for each 
event.  It is probable that 
the atmosphere deviates from the standard conditions from night to 
night, a view 
strengthened by the results of balloon flights made from 
Malarg\"{u}e \cite{bal03}. These have shown that the 
atmosphere changes in a significant way from night to night, and 
from summer to winter.  If a standard atmosphere is used, 
some of the fluctuations observed in X$_{max}$ may be incorrectly 
attributed to shower, rather than to atmospheric, variations.  
Thus, it may be premature to draw conclusions about the 
presence of protons from this, and similar earlier analyses.

\subsection{ Mass from muon density measurements} 
\begin{figure}[htb]
\centerline{\epsfxsize=5.0in\epsfbox{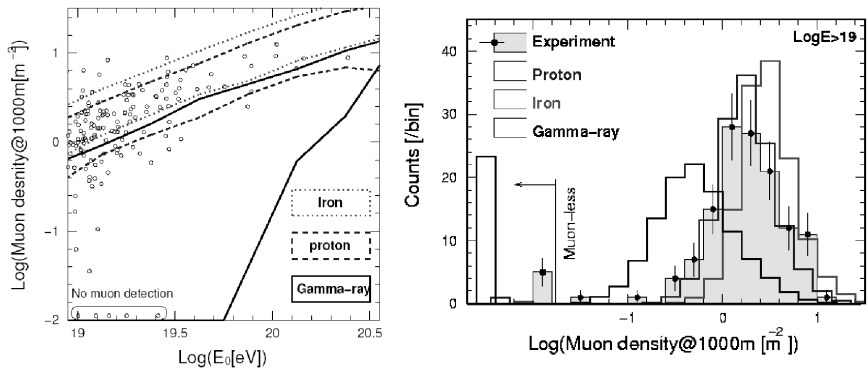}}
\caption{The muon density at 1000 m as measured at AGASA [13].  
In the left hand diagram, 
the dotted lines are the predictions for iron nuclei, the dashed 
lines for protons and the solid lines for 
photons.  In the right hand diagram, the shaded histogram represents 
the data with predictions for iron, 
protons and photons shown by the line histograms: iron is the 
right-most histogram. \label{fig4}}
\end{figure}

A shower produced by an iron 
nucleus will contain a larger fraction of muons at the observation 
level than a shower of the same energy created by a proton.  
Many efforts to uncover the mass spectrum of cosmic rays have 
attempted to use this fact.  However, although the differences 
are predicted to be 
large ($\sim$70\% more muons in an iron event than a proton event), 
there are large fluctuations 
and, again, there are differences between what is predicted by 
particular models.  Thus, more muons are predicted with the QGSJET 
set than with the Sibyll family.  The difference 
arise from different predictions as to 
the pion multiplicities in nucleon-nucleus and pion-nucleus 
collisions \cite{alv02}.  Recent  
data from the AGASA group \cite{amd03} is shown in figure~\ref{fig4}.  There 
are 129 events above $10^{19}$ eV, of which 
19 have energies $>$ $3 \times 10^{19}$ eV.  Measurements 
at distances between 800 
and 1600 m were used to derive the muon density at 1000 m with 
an average accuracy of 40\%.  These muon densities 
are compared with the predictions of model calculations.  It 
is clear that the 
difference between the proton and iron predictions is small, 
especially when fluctuations are 
considered.  The AGASA group conclude that at $10^{19}$ eV the 
fraction of Fe nuclei is $14^{+16}_{-14}$\% and 
above $3 \times 10^{19}$ eV it 
is $30^{+7}_{-6}$\%.  Of the 5 events above $10^{20}$ eV, 
3 are as well fitted by 
iron nuclei as by protons.
%

The conclusions are sensitive to the model used and as the 
Sibyll model predicts fewer muons than the QGSJET model, higher 
iron fractions would have been inferred had that model been adopted.  

At lower energies, there are muon data from the Akeno array and 
from AGASA \cite{hay95}.  Different 
analyses have been made.  The AGASA group\cite{hay95}  claim 
that the measurements are consistent with a mass composition that 
is unchanging between $10^{18}$ and $10^{19}$ eV.   Dawson 
et al.\ \cite{dawms}, in an effort to reconcile these 
data with earlier fluorescence results have argued 
that, with a different model, the mean mass is lower at 
higher energies.  In the context of the present 
discussion, it is worth noting that 50 -- 60\% of iron,
near $10^{19}$ eV, is quite 
consistent with both the AGASA and 
Akeno data for a range of models and with efforts to account 
for systematic uncertainties.  It might be 
productive to re-examine these data using the latest QGSJET 
and Sibyll models.

\subsection{Mass estimates from the lateral distribution function}  
The rate of fall of particle density with distance from the 
shower axis provides another parameter that can be used to 
extract the mass composition.  Showers with steeper 
lateral distribution functions (LDFs) than average will arise from 
showers that develop later in the atmosphere, and vice versa.  A 
detailed measurement of the LDFs of 
showers produced by primaries of energy greater 
than $10^{17}$ eV was made at Haverah Park using a 
specially constructed `infilled array' in which 30 additional water 
tanks of 1 m$^{2}$ were added  on a grid with spacing of 150 m.  
When the work was 
completed in 1978, the data could not be fitted with the 
shower models then available.  Recently \cite{ave02}, the data 
have been re-examined using the QGSjet98 model.

\begin{figure}[htb]
\centerline{\epsfxsize=4.5in\epsfbox{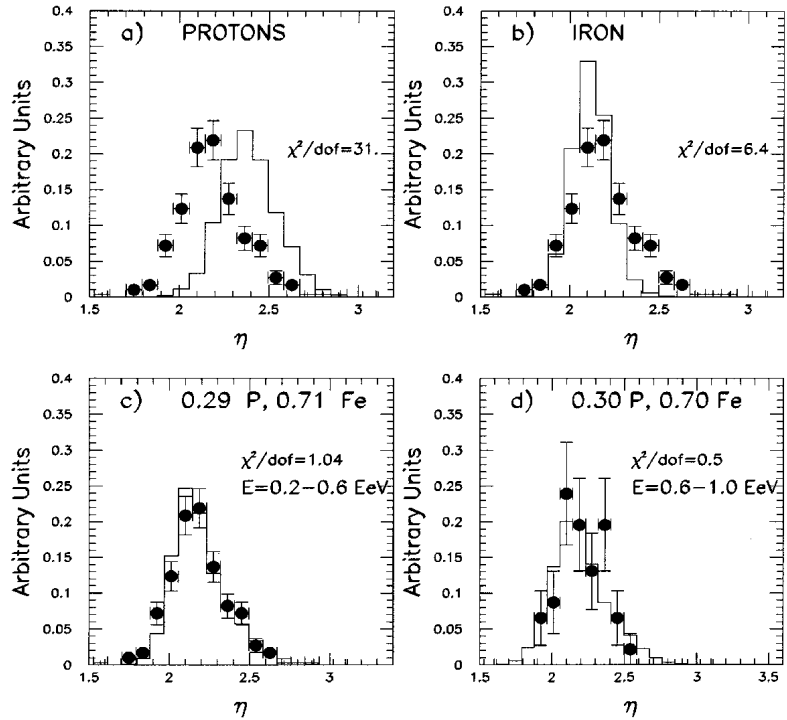}}
\caption{The experimental measurements of the steepness 
parameter, $\eta$, compared with predictions made 
using the QGSJET98 model assuming different mass mixtures $^{16}$.  
The lower set of diagrams illustrates the 
insensitivity of the mass mixture to energy. \label{fig5}}
\end{figure}

The appropriateness of this model was established by showing that 
it adequately described data on the 
time spread of the Haverah Park detector signal over a range of 
zenith angles and distances near the core 
($<$500 m).  Here the difference predicted between the average proton 
and iron shower is only a few 
nanoseconds and the 
fit is good.  Densities were fitted by 
$\rho$(r)$\sim$r$^{-(\eta + r/4000)}$, where 
$\eta$ is the steepness parameter.  The spread of $\eta$ is 
compared with predictions  
in figure~\ref{fig5}.  The proton fraction, assuming a 
proton-iron mixture, is found to be independent of energy 
in the range $3 \times 10^{17}$ to $10^{18}$ eV and is (34 $\pm$ 2) \%.  
If this is evaluated with QGSJET01, in which 
a different treatment of diffractive processes is adopted,  then the 
fraction increases to 48\%.  It is larger 
because the later model 
predicts shower maxima that are higher 
in the atmosphere and accordingly, to match the observed fluctuations, the proton fraction must be increased.  The 
deduced ratio thus has a systematic uncertainty from the models that is 
larger than the statistical uncertainty.  Although the necessary 
analysis has not been made, it is clear that 
the Sibyll 2.1 model would require a smaller fraction of protons.  
%

A similar analysis has been carried out using data from the Volcano 
Ranch array.  As with the Haverah Park information, no 
satisfactory interpretative analysis was possible when the measurements 
were made.  With QGSJET01, the fraction of protons is estimated 
as (25 $\pm$ 5\%) between 5 and $10 \times 10^{18}$ eV \cite{dov03}. 

\subsection{Mass from the thickness of the shower disk} 
The particles in the shower disc do not arrive at a 
detector simultaneously.  The arrival times 
are spread out because of geometrical effects, velocity 
differences, multiple scattering and 
geomagnetic deflections.  The first particles to arrive (except very 
close to the shower axis) are the muons: they are scattered 
little and geometrical effects dominate.  At Haverah Park four 
detectors, each of 34 m$^{2}$, provided a useful tool for studying the 
thickness of the shower disc, which depends upon the development 
of the cascade.  Recently, an analysis of 100 events has shown that the magnitude of 
the risetime is indicative of a large fraction ($\sim$80\%) iron 
nuclei at $\sim10^{19}$eV \cite{ave03}.  This type 
of study will be considerably extended with the 
Pierre Auger Observatory, in which each water tank is equipped 
with 25 ns flash ADCs \cite{dovmt}.

\subsection{Limits to the fraction of photon primaries}  
It is unlikely that the majority of the events claimed 
to be near $10^{20}$ eV have photons as parents as some 
of the showers seem to have normal numbers of muons (the 
tracers of primaries that are baryonic), figure~\ref{fig4}.  Furthermore, 
the cascade profile of the most 
energetic fluorescence \cite{bir93} event is inconsistent with that 
of a photon \cite{hal95}. This approach can be 
used when specific shower profiles are available.  An alternative 
method of searching for photons has 
recently been developed using showers incident at very large 
zenith angles.  Deep-water tanks have a 
good response to such events out to beyond $80^{\circ}$.  At such 
angles the bulk of the showers detected are 
created by baryonic primaries but are distinctive in that the 
electromagnetic cascade stemming from neutral pions has 
been completely suppressed by the extra 
atmosphere.  At $80^{\circ}$ the atmospheric thickness 
to be penetrated is $\sim$ 5.7 atmospheres.  At Haverah Park, such 
large zenith angle showers were observed and 
the shower disc was found to have a very small 
time spread.  A complication for the study of inclined showers is that 
the muons, in their long traversal 
of the atmosphere, are significantly bent by the geomagnetic 
field.  A detailed study of this has been made and it 
has been shown that the rate of triggering of the Haverah Park 
array at large angles can be predicted \cite{ave00}.  In addition, it 
was shown that the energy of the primaries could be estimated 
with reasonable precision and an energy spectrum 
derived.  The concept of using the 
known, and mass independent, spectrum deduced by the fluorescence 
detectors to predict the 
triggering rate as a function of the mass of the primary has led to a demonstration that the photon flux 
at $10^{19}$ eV is less than 40\% of the baryonic 
component \cite{avm00}.  In addition to this novel 
approach, a more traditional attack on the 
problem by the AGASA group, searching for showers which have 
significantly fewer muons than 
normal, has given the same result \cite{shi02}.  These experimental 
limits are in contrast to the predictions 
of large photon fluxes from the decay of super-heavy relic 
particles, one of the exotic candidates that 
have been invoked to explain the enigma \cite{ber97}.

A first attempt at estimating the photon flux at $10^{20}$ eV 
can be made from the data of figure~\ref{fig4} and other 
observations.  In addition to the Fly's Eye event discussed 
in \cite{ave03}, the HiRes group has reported one 
stereo event \cite{loh02} that is nearly as large and with a 
longitudinal development profile that does not 
match that of a photon.  At least one of the very inclined events 
in the study of \cite{avm00} is above $10^{20}$ eV 
and the Yakutsk event \cite{efi90} of $57^{\circ}$ is very rich in 
muons.  There are thus 9 events above $10^{20}$ eV for 
which a judgement about their photonic nature can be made.  If none 
of the 5 AGASA events above $10^{20}$ 
eV shown in figure ~\ref{fig4} is a photon, then the 95\% confidence 
limit for the photon flux, as a fraction of 
the total cosmic ray flux, is 33\%.  If two of the AGASA events 
are produced by photons, then the flux 
is estimated as ($22^{+30}_{-14}$\%).  Both figures are of interest 
in the context of the super-heavy relic 
models \cite{ber97}.  Further details will be given 
elsewhere \cite{dovwa}.  

\section*{Conclusions} 
The question of spectral shape of the UHECRs remains uncertain 
and, along with the issue of the clustering 
of the arrival directions, may only be resolved by the operation 
of the Pierre Auger Observatory.  To make full 
use of this forthcoming information, it is necessary to improve our 
knowledge of the mass of the cosmic rays above $10^{19}$ eV.  Such 
evidence as there is does not support 
the common assumption that all of these cosmic rays are 
protons: there may be a substantial fraction 
of iron nuclei present.  Photons do not appear to dominate 
at the highest energies. 

\section*{Acknowledgements}  
I would like to thank the organisers 
for inviting me to the Sorrento meeting and, 
in particular, Michelangelo Ambrosio for his hospitality.  Work 
on UHECR at the University of Leeds is supported by PPARC, UK.

\end{document}